\documentclass[pdflatex,sn-mathphys-num]{sn-jnl}% Math and Physical Sciences Numbered Reference Style 
\usepackage{graphicx}%
\usepackage{multirow}%
\usepackage{amsmath,amssymb,amsfonts}%
\usepackage{amsthm}%
\usepackage{mathrsfs}%
\usepackage[title]{appendix}%
\usepackage{xcolor}%
\usepackage{textcomp}%
\usepackage{manyfoot}%
\usepackage{booktabs}%
\usepackage{algorithm}%
\usepackage{algorithmicx}%
\usepackage{algpseudocode}%
\usepackage{listings}%
\usepackage{siunitx}%additionally added
\usepackage{subcaption}%additionally added
\theoremstyle{thmstyleone}%
\theoremstyle{thmstyletwo}%

\theoremstyle{thmstylethree}%

\begin{document}

\title[Generating 3D Pseudo-Healthy Knee-MR Images to Support Trochleoplasty Planning]{Generating 3D Pseudo-Healthy Knee MR Images to Support Trochleoplasty Planning}


\author*[1]{\fnm{Michael} \sur{Wehrli}}\email{michaeljan.wehrli@unibas.ch}

%%\author[2,3]{\fnm{Second} \sur{Author}}\email{iiauthor@gmail.com}

\author[1]{\fnm{Alicia} \sur{Durrer}}%\email{@unibas.ch}

\author[1]{\fnm{Paul} \sur{Friedrich}}%\email{@unibas.ch}

\author[1]{\fnm{Volodimir} \sur{Buchakchiyskiy}}%\email{@unibas.ch}

\author[2]{\fnm{Marcus} \sur{Mumme}}%\email{@unibas.ch}

\author[2]{\fnm{Edwin} \sur{Li}}%\email{@unibas.ch}

\author[2]{\fnm{Gyozo} \sur{Lehoczky}}%\email{@unibas.ch}

\author[2]{\fnm{Carol C.} \sur{Hasler}}%\email{@unibas.ch}

\author[1]{\fnm{Philippe C.} \sur{Cattin}}%\email{philippe.cattin@unibas.ch}

\affil[1]{\orgdiv{Department of Biomedical Engineering}, \orgname{University of Basel}, \city{Allschwil},  \country{Switzerland}}

\affil[2]{\orgdiv{Department of Orthopedics}, \orgname{University Children’s Hospital Basel} \country{Switzerland}}

%\affil[1]{\orgdiv{Department of Biomedical Engineering}, \orgname{University of Basel}, \orgaddress{\street{Hegenheimermattweg 167A}, \city{Allschwil}, \postcode{4123}, \state{Baselland}, \country{Switzerland}}}

%\affil[2]{\orgdiv{Department of Orthopedics}, \orgname{University Children’s Hospital Basel}, \orgaddress{\street{Spitalstrasse 33}, \city{Basel}, \postcode{4031}, \state{Basel}, \country{Switzerland}}}

%%\affil[3]{\orgdiv{Department}, \orgname{Organization}, \orgaddress{\street{Street}, \city{City}, \postcode{610101}, \state{State}, \country{Country}}}


\abstract{\textbf{Purpose:} Trochlear Dysplasia (TD) is a common malformation in adolescents, leading to anterior knee pain and instability. Surgical interventions such as trochleoplasty require precise planning to correct the trochlear groove. However, no standardized preoperative plan exists to guide surgeons in reshaping the femur. This study aims to generate patient-specific, pseudo-healthy MR images of the trochlear region that should theoretically align with the respective patient’s patella, potentially supporting the pre-operative planning of trochleoplasty.

\textbf{Methods:} We employ a Wavelet Diffusion Model (WDM) to generate personalized pseudo-healthy, anatomically plausible MR scans of the trochlear region. We train our model using knee MR scans of healthy subjects. During inference, we mask out pathological regions around the patella in scans of patients affected by TD, and replace them with their pseudo-healthy counterpart. An orthopedic surgeon measured the sulcus angle (SA), trochlear groove depth (TGD) and Déjour classification in MR scans before and after inpainting. The code is available at \url{https://github.com/wehrlimi/Generate-Pseudo-Healthy-Knee-MRI}.

\textbf{Results:} The inpainting by our model significantly improves the SA, TGD and Déjour classification in a study with 49 knee MR scans.

\textbf{Conclusion:} This study demonstrates the potential of WDMs in providing surgeons with patient-specific guidance. By offering anatomically plausible MR scans, the method could potentially enhance the precision and preoperative planning of trochleoplasty, and pave the way to more minimally invasive surgeries.
}




\maketitle

\section{Introduction}\label{sec1}

\begin{figure}[H]
    \centering
    \includegraphics[width=0.9\textwidth]{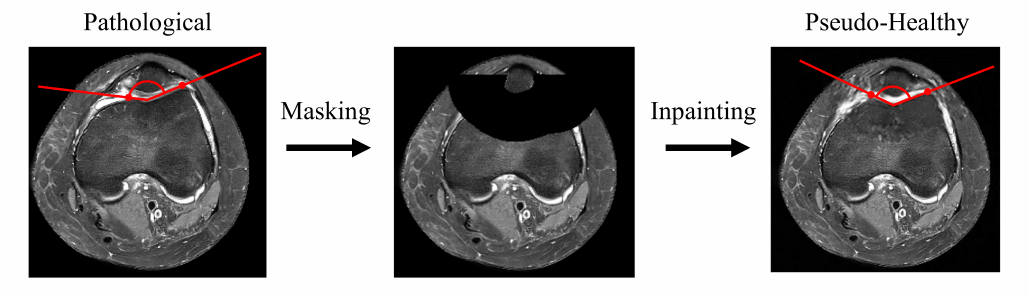}
    \caption{Axial slices of an MR scan showing: (left) a pathological knee with TD indicated by an abnormally large sulcus angle (visualized in red), (middle) a masked image and (right) a pseudo-healthy knee with a normal sulcus angle. We work with 3D volumes, but for visualization purposes, only 2D slices are shown here.\label{fig_TD}}
\end{figure}

Trochlear Dysplasia (TD) is an anatomical deformity that demands precise corrective interventions in order to address anterior knee pain and instability in adolescents \cite{batailler_trochlear_2018}. This condition, characterized by a malformation of the trochlea femoris, results in lateralized tracking of the patella, thereby elevating the risk of dislocation \cite{bollier_role_2011, liebensteiner_diagnostik_2023, dirisamer_klinische_2023, wierer_therapiealgorithmus_2023}. An exemplary case is shown in Figure~\ref{fig_TD} on the left. In addition to the discomfort and functional limitations it imposes, untreated TD can lead to the development of osteoarthritis, resulting in chronic pain and impaired mobility \cite{hasler_patella_2016, fithian_epidemiology_2004}.
Despite the widespread adoption of trochleoplasty, the surgery to correct this malformation, many patients report continued pain and have variable perceptions of satisfaction \cite{blond_trochlear_2023}. One of the key challenges in this surgery is achieving the correct trochlear shape without damaging the fragile femoral cartilage, a task complicated by individual anatomical variations \cite{blond_arthroscopic_2015}. While various diagnostic methods exist, the Déjour criterion is the most commonly used in clinical practice \cite{dejour_factors_1994, biedert_patellotrochlear_2006}. These methods offer insights into the diagnosis and categorization of TD without generating a patient-specific plan for its correction. This diagnostic-centric approach inadvertently creates a gap between the identification of anatomical anomalies and the execution of precise, reproducible surgical manipulations. Currently, surgical outcomes rely heavily on the subjective image assessment of surgeons, who must adapt the trochlear shape to match the patella, a task complicated by the natural variation in patella shapes. Since each patella requires a corresponding trochlear shape, individualized preoperative planning is crucial. Due to these difficulties, most trochleoplasty procedures are open surgeries, which involve larger incisions and longer recovery times \cite{batailler_trochlear_2018, hinckel_trochleoplasty_2015, nolan_trochleoplasty_2018}. Therefore, it is necessary to re-evaluate and enhance current procedures \cite{nolan_trochleoplasty_2018, knoch_trochleaplasty_2006}. 
 
A recent study by Barbosa et al. \cite{barbosa_knee_2024} introduced a deep learning approach to detect knee landmarks for TD. But, a detailed preoperative patient-specific plan, e.g., in Fang et al. \cite{fang_patient-specific_2024} for orthognathic surgery, has not yet been established. While statistical shape modeling has been used to visualize abnormalities in trochleas \cite{van_haver_statistical_2014}, it has not yet provided a clear corrective plan. In addition, these models do not consider individual patella anatomy and require CT imaging that exposes patients to harmful radiation. In contrast, our approach uses preoperative MR images, which are already part of standard patient care. No additional CT images are necessary for our approach, reducing the radiation dose for these mainly young patients. Therefore, we propose to generate pseudo-healthy MR images to support the planning process of this complicated surgery \cite{beaufils_trochleoplasty_2012, hinckel_trochleoplasty_2015, nolan_trochleoplasty_2018}. This could potentially improve the treatment outcome for the patients, paving the way to a more minimally invasive approach by supporting the surgeon in their preoperative decision. In the future, pseudo-healthy MR images could also be used as a preoperative plan to guide the surgeon in the operation room, reducing the necessity for open surgery.

%\newpage
\section{Methods}\label{methods}

Our proposed method generates pseudo-healthy MR images, restoring the trochlear shape while being conditioned on the patient's unique patella. Following \cite{durrer_denoising_2024}, we mask pathological regions and employ a Wavelet Diffusion Model by \cite{friedrich_wdm_2024} for inpainting healthy tissue.

\subsection{Foreground-Background Segmentation and Masking}
\begin{figure}[H]
\centering
\includegraphics[width=0.9\textwidth]{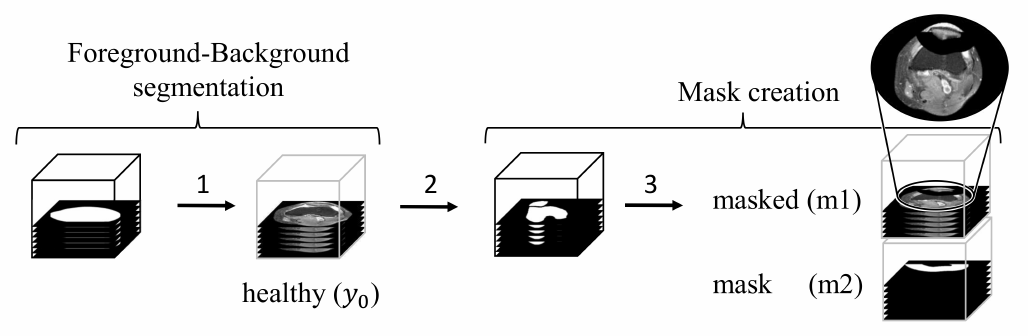}
\caption{After we identified the background region, we set it to zero and get our healthy target image $y_0$ as shown in Step 1. In Step 2, we segment the bone. In Step 3 and based on this segmentation, we create a mask around the patella.\label{Fig_Masking}}
\end{figure}

As we treat the problem as an inpainting task, we first need to identify pathological regions and mask them. In the first step, we get rid of any background noise that hinders the performance of the WDM; see Figure~\ref{Fig_Masking} Step 1. By using Otsu thresholding, followed by morphological opening and closing, and then extraction of the largest connected component, we segment the background. Afterward, we segment the bones using SegmentAnyBone \cite{gu_segmentanybone_2024}, compare to Step 2. Based on this segmentation, we localize the patella. After patella localization, we then create a mask around it with a \SI{30}{\milli\metre} offset, forming a "bowl" around the patella, see Step 3. 

In cases where the patella is not found due to segmentation errors, a 3D ellipsoid mask is used instead. 
 
\subsection{Background about Denoising Diffusion Probabilistic Models}
Denoising Diffusion Probabilistic Models (DDPMs) are generative models consisting of an iterative forward process $q$ and a learned reverse process $p_{\theta}$. The forward process $q$ adds noise to an input image $x_0$ for $T$ time steps $t$. It can also be viewed as a Markov chain, where each transition follows a Gaussian distribution

\begin{equation}
q(x_t | x_{t-1}) = \mathcal{N}(x_t; \sqrt{1-\beta_t}x_{t-1}, \beta_t \mathbf{I}) 
\end{equation}

with the identity matrix $\mathbf{I}$ and the fixed forward process variances $\beta_1, \dots, \beta_T$. This noising process transforms the initial image \(x_0\) into a standard normal distribution $x_T$ for sufficiently large $T$. The reverse process $p_{\theta}(x_{t-1} | x_t)$ attempts to remove the added noise and follows a sequence of Gaussian distributions parameterized by a neural network

\begin{equation}
p_{\theta}(x_{t-1} | x_t) = \mathcal{N}(x_{t-1}; \mu_{\theta}(x_t, t), \Sigma_{\theta}(x_t, t))
\end{equation}

where $\mu_{\theta}$ and $\Sigma_{\theta}$ represent the mean and variance and are computed by a neural network $\epsilon_{\theta}(x_t, t)$.

\subsection{Wavelet Transform}
The Discrete Wavelet Transform (DWT) is a tool in time-frequency analysis, commonly used to decompose images into multiple frequency bands and has also been used for diffusion models in \cite{friedrich_wdm_2024}. This allows for efficient processing of both low- and high-frequency components. The DWT decomposes a volume into 8 wavelet coefficients: $ (x_{\text{lll}}, x_{\text{llh}}, x_{\text{lhl}}, x_{\text{lhh}}, x_{\text{hll}}, x_{\text{hlh}}, x_{\text{hhl}}, x_{\text{hhh}}) = \text{DWT}(y_0) = x_0 $, also visible in Figure~\ref{fig_training_details}. Each wavelet coefficient has half the spatial resolution $ x_{\text{lll}, \dots, \text{hhh}} \in \mathbb{R}^{\frac{D}{2} \times \frac{H}{2} \times \frac{W}{2}} $. The low-frequency component $x_{\text{lll}}$ is akin to a downsampled version of the image, while the other components $x_{\text{llh}, \dots, x_{\text{hhh}}}$ capture high-frequency details such as edges and textures. To reconstruct the original image $y_0$, the Inverse Discrete Wavelet Transform (IDWT): $y_0 = \text{IDWT}(x_{\text{lll}}, \dots, x_{\text{hhh}})$ is applied. In this work, as in \cite{friedrich_wdm_2024}, we use the Haar wavelet transform implementation from Li et al. \cite{li_wavelet_2020}.

\subsection{Modifications of DDMPs for the Inpainting Training Procedure}

\begin{figure}[H]
\centering
\includegraphics[width=\textwidth]{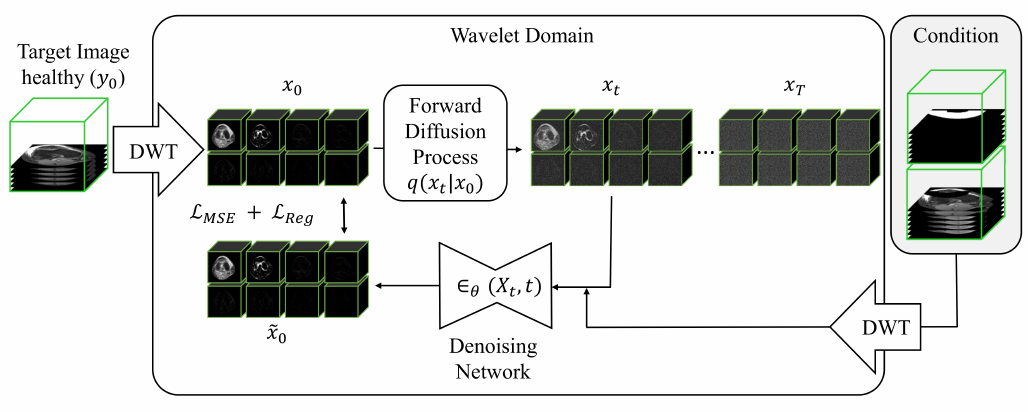}
\caption{\textbf{Training on healthy MR scans:} During training, the noisy wavelet decomposed image $x_t$ is concatenated together with the condition (the DWT of the masked image $m_1$ and mask $m_2$). During training, the network is tasked with predicting the denoised image $\tilde{x}_0$ based on $x_t$ at a random time step $t$. The loss is then calculated between the prediction $\tilde{x}_0$ and $x_0$. \label{fig_training_details}}

\end{figure}

To adapt DDPMs for the inpainting task, we modify the reverse process by conditioning $x_t$ with a concatenation of the DWT of the masked image $m_1$ and the mask $m_2$, see also Figure~\ref{fig_training_details}. Specifically, the modified input at each timestep becomes: $ X_t := (x_t, DWT(m_1), DWT(m_2)) $. This concatenated input is fed into the neural network $\epsilon_{\theta}(X_t, t)$, as shown in Figure~\ref{fig_training_details}, which conditions the denoising process. The model learns to reconstruct $x_0$ within the wavelet domain (called $\tilde{x}_0$). If we would apply an IDWT to $\tilde{x}_0$, we would go back to the original representation domain and have our inpainted image $\tilde{y}_0$.

During training, the network is tasked with predicting the denoised image $\tilde{x}_0$ based on $x_t$ at a random time step $t$. The noisy image $x_t$ is generated as $x_t = \sqrt{\bar{\alpha}_t}x_0 + \sqrt{1 - \bar{\alpha}_t} \epsilon, \quad \text{where} \quad \epsilon \sim \mathcal{N}(0, \mathbf{I})$ with $\alpha_t = 1 - \beta_t$ and $\bar{\alpha}_t = \prod_{s=1}^{t} \alpha_s$. The model is trained using the Mean Squared Error (MSE) loss between the predicted denoised image $\tilde{x}_0$ and the ground truth $x_0$. Additionally, a L1 Regularization term is applied to the high-frequency (HF) wavelet components to encourage sparsity and reduce HF noise around edges:

\begin{equation}
\mathcal{L} = \mathcal{L}_{\text{MSE}} + \mathcal{L}_{\text{Reg}} = \| \tilde{x}_0 - x_0 \|_2^2 + \left( \| x_{\text{hhh}} \|_1 + \| x_{\text{hhl}} \|_1 + \| x_{\text{hlh}} \|_1 + \| x_{\text{lhh}} \|_1 \right)
\end{equation}

\subsection{Inference Procedure}

\begin{figure}[H]
\centering
\includegraphics[width=\textwidth]{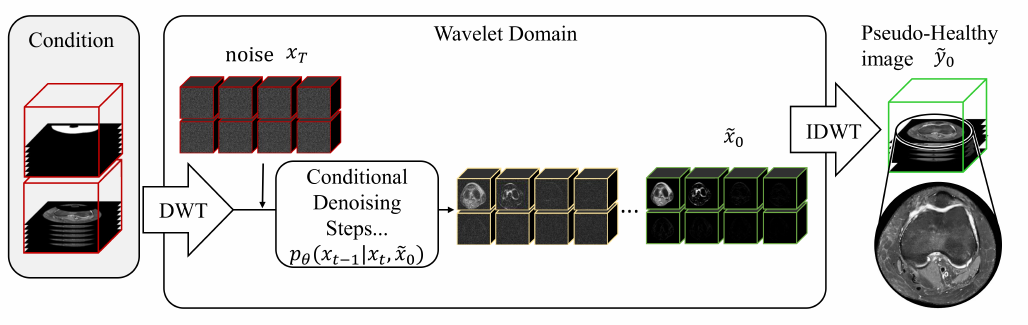}
\caption{\textbf{Inference:} Starting from the left side with a masked pathological knee MR scan (shown in red) forming our condition. This condition ($m_1$ and $m_2$) is now used to iteratively remove noise from the noisy image $x_T$ using a reverse process. After $T$ steps, we perform the IDWT to obtain a pseudo-healthy image (shown in green).\label{fig_inference}}

\end{figure}

To sample from the trained model, we compute

\begin{equation}
\mu_t(\mathbf{x}_t, \tilde{\mathbf{x}}_0) := \frac{\sqrt{\bar{\alpha}_t - 1}\beta_t}{1 - \bar{\alpha}_t} \tilde{\mathbf{x}}_0 + \frac{\sqrt{\alpha_t} (1 - \bar{\alpha}_{t-1})}{1 - \bar{\alpha}_t} \mathbf{x}_t,
\end{equation}
\begin{equation}
\tilde{\beta}_t := \frac{1 - \bar{\alpha}_{t-1}}{1 - \bar{\alpha}_t} \beta_t.
\end{equation}
defining the following distribution:
\begin{equation}\label{denoising_equation}
p_\theta(\mathbf{x}_{t-1} | \mathbf{x}_t, \tilde{\mathbf{x}}_0) = \mathcal{N} \left( \mu_t(\mathbf{x}_t, \tilde{\mathbf{x}}_0), \tilde{\beta}_t \mathbf{I} \right)
\end{equation}
We draw from Equation~\ref{denoising_equation} at each denoising step to get $x_{t-1}$. This process is performed in the wavelet domain, and the final denoised image is obtained by applying an inverse discrete wavelet transform (IDWT). The key idea is to use this inference procedure on images affected by TD, see Figure~\ref{fig_inference}, and generate a pseudo-healthy MR image for this specific patient.

\section{Experiments}

\subsection{Data}

For training, we use healthy knee MR data from the publicly available fastMRI dataset (\url{https://fastmri.med.nyu.edu/}) \cite{noauthor_fastmri_nodate, zbontar_fastmri_2019}, containing 10'012 knee images, forming a representative clinical patient population. We use proton-density weighted (PD) and fat-saturated (FS or F/S) images only, as these are most frequently used to assess TD. Specifically, we chose the ones with "PD AXIAL F/S" or "PD AXIAL FAT SAT" in their series description. Out of 10'012 knee volumes, 1'579 sequences matched these criteria. Out of these, we used 1'518 DICOM volumes (1'477 unique patients) with constant slice thickness. We split these into training (80\%, 1'181 unique patients, 1'216 volumes) and test (20\%, 296 unique patients, 302 volumes) sets. Additionally, we use in-house MR images with TD from the Children's University Hospital Basel to evaluate our model and workflow on pathological data. From the available 117 images (52 patients) we excluded 68 images (20 patients), containing dislocated patellae, the presence of extreme swelling or prominent growth plates.

\subsection{Ethical approval declaration}
The dataset from the University Children's Hospital Basel was anonymized and granted exemption by the Ethikkommission Nordwest- und Zentralschweiz (Req-2024-01188). The methods were carried out in accordance with the relevant guidelines and regulations. 

\subsection{Preprocessing}
\begin{figure}[H]
\centering
\includegraphics[width=0.9\textwidth]{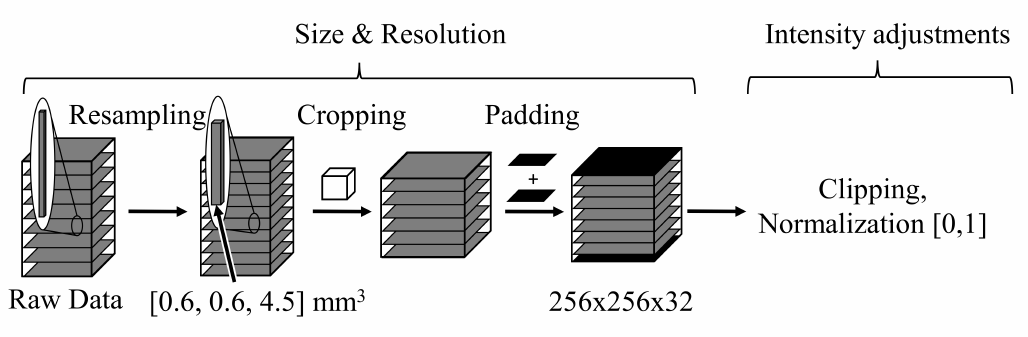}
\caption{First, we resample all volumes to the voxel dimension of $(0.6 \text{mm}, 0.6 \text{mm}, 4.5 \text{mm})$. In the second step, we crop from the center up to $(256, 256, 32)$ voxels or the end of the current volume. Next, we pad the volume if necessary to reach the size $(256, 256, 32)$ voxels. Lastly, we adjust the intensity range.\label{Fig_Preprocessing}}
\end{figure}

These PD fat-saturated sequences that are most frequently used to assess TD exhibit a higher resolution in the $xy$-plane compared to the $z$-direction. The voxel size of the axial PD fat-saturated fastMRI knee scans is $(0.60 \pm 0.03\,\text{mm}, 0.60 \pm 0.03\,\text{mm}, 4.43 \pm 0.38\,\text{mm})$. The mean shape of these images is $(256.52 \pm 13.29, 256.63 \pm 13.18, 29.31 \pm 3.56)$, with a median shape of $(256, 256, 29)$ voxels. To standardize the image size across all volumes, we preprocess the data to a fixed size of $(256, 256, 32)$ voxels. The same preprocessing steps are applied to both the fastMRI dataset and the in-house dataset: First, we resample the voxel size to the same dimensions $(0.6 \text{mm}, 0.6 \text{mm}, 4.5 \text{mm})$. If the images are too large, we first center-crop them, then pad the smaller volumes where necessary, ensuring all images reach a size of $(256, 256, 32)$ voxels. Lastly, we clip the highest and lowest percentile and normalize between [0,1], see also Figure~\ref{Fig_Preprocessing}.

\subsection{Diagnostic evaluation criteria}
We compared 49 knee MR scans of 32 patients with TD before and after inpainting based on known metrics to assess TD. These are sulcus angle (SA), trochlear groove depth (TGD) and Déjour classification, see Figure~\ref{fig_TD}. The measurements were performed by a deputy attending orthopedic surgeon using 3D Slicer \cite{kikinis_3d_2014, noauthor_3d_nodate} Version 5.6.2.

\subsection{Image quality evaluation metrics}
We evaluated the generated inpainting in the fastMRI test set images in terms of Mean Squared Error (MSE), Peak Signal-to-Noise Ratio (PSNR) and Structural Similarity Index Measure (SSIM). All scores were calculated over the masked regions of the image only. MSE and PSNR measure pixel-wise accuracy, with lower MSE and higher PSNR indicating better fidelity. SSIM assesses structural and perceptual quality, making it crucial for evaluating visual similarity. Higher SSIM values indicate better visual similarity and perceptual quality. For SSIM, we first zero all non-mask voxels, then apply regular SSIM on the complete volume. In the end, we take the "full SSIM" image from torchmetrics and only take the values relating to voxels within the mask.

\subsection{Implementation Details}
The hyperparameter settings are described in Table~\ref{table:hyper}. Except for the resolution, they are identical to Friedrich et al. \cite{friedrich_wdm_2024}.  The used model has 57'942'472 parameters. All experiments were carried out on a single NVIDIA A100 (40 GB) GPU, with an average inference runtime of 45\,s and a memory footprint of 3.81\,GB.

\begin{table}[h]
\centering
\caption{Hyperparameter settings \label{table:hyper}}
\begin{tabular}{|l|l|l|}
\hline
Dataset                       & fastMRI         & \begin{tabular}[c]{@{}l@{}}In-house dataset \\ (Inference only)\end{tabular} \\ \hline
Resolution                    & $256\times256\times32$      & $256\times256\times32$                                                                   \\ \hline
Number of base channels        & 64              & 64                                                                           \\ \hline
Number of ResBlocks per scale  & 2               & 2                                                                            \\ \hline
Channel multiplier per scale   & (1, 2, 2, 4, 4) & (1, 2, 2, 4, 4)                                                              \\ \hline
Learning rate                  & $1 \times 10^{-5}$ & $1 \times 10^{-5}$                                                          \\ \hline
Batch size                     & 10              & 1                                                                            \\ \hline
Number of training iterations  & $1.0 \times 10^{6}$ & N/A                                                                      \\ \hline
Diffusion timesteps T          & 1000            & 1000                                                                         \\ \hline
Noise schedule                 & linear          & linear                                                                       \\ \hline
\end{tabular}
\end{table}

%\newpage
\section{Results}\label{sec3}
\subsection{Diagnostic evaluation}

\begin{figure}[h]
\centering
\includegraphics[width=1\textwidth]{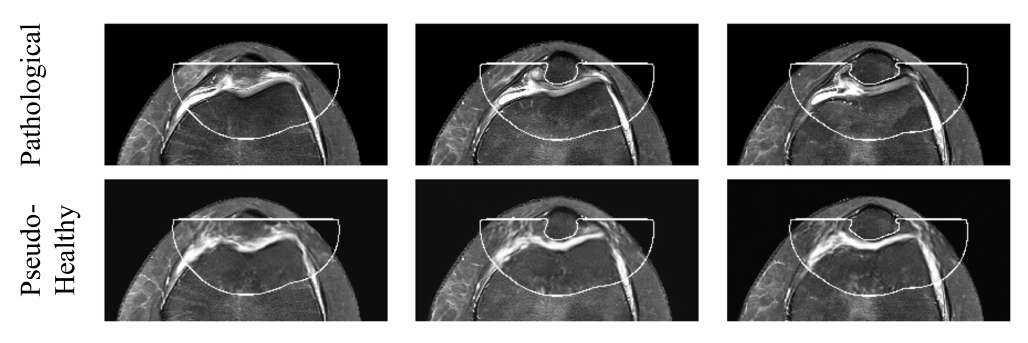}
\caption{An example image. Top row: Consecutive slices from a patient's pathological MR scan prior to inpainting. Bottom-row: The same slices after inpainting with the WDM, showing a pseudo-healthy version. The white lines mark the edges of the mask. We work with 3D volumes, but for visualization purposes, only 2D slices are shown.\label{fig:results_qualitative}}
\end{figure}
\begin{figure}[h]
\centering
\includegraphics[width=1\textwidth]{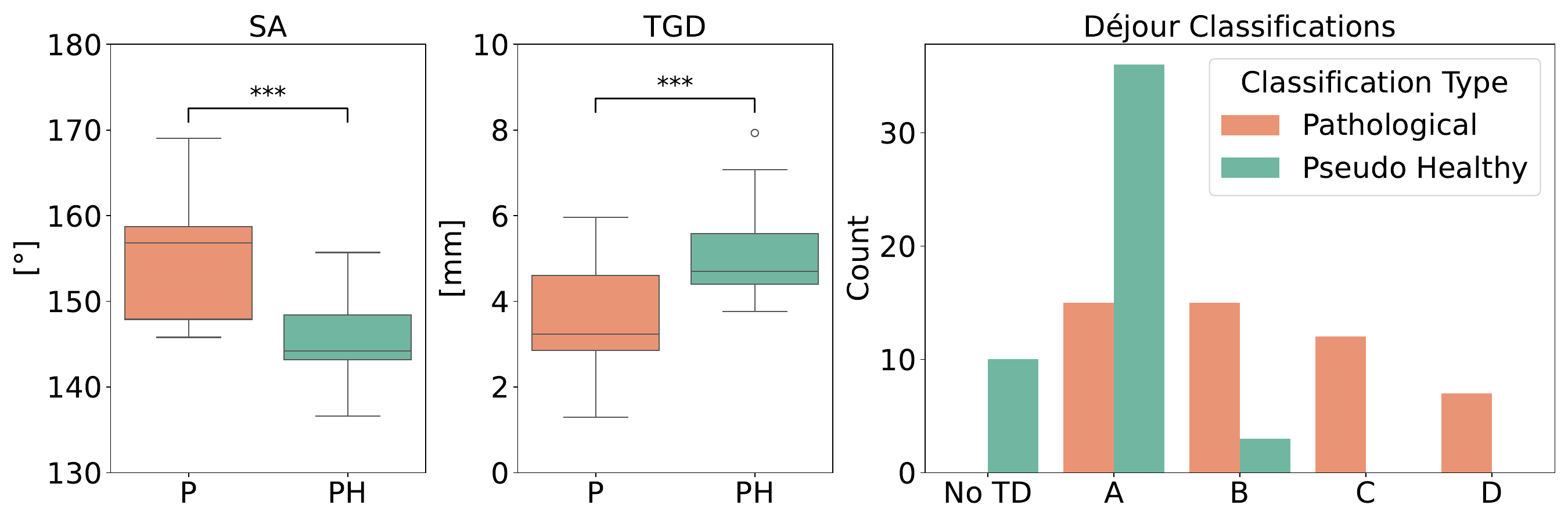}
\caption{Left: The difference between the SA in the 16 scans is significant by the Wilcoxon signed-rank test. Middle: Significant difference in TGD. Right: A reduction in severity in the pseudo-healthy versions is evident. \label{fig:results_measurements}}
\end{figure}

 
Out of the 49 pathological MR scans, 41 showed a reduction in the severity of TD, as classified by Déjour's method, by one or more stages. Figure~\ref{fig:results_measurements} (right) demonstrates this shift, showing a clear reduction in severity in the pseudo-healthy versions.

For 33 out of 49 scans, SA and TGD could not be measured before inpainting due to the severity of the TD. After inpainting, however, the pseudo-healthy scans allowed for measurable SA and TGD in 46 out of 49 cases.
Among the 16 scans that had a measurable SA and TGD both before and after inpainting, quantitativ analysis revealed a significant reduction in SA (from a pathological mean of $154^\circ$ to a pseudo-healthy mean of $145^\circ$), as shown in Figure ~\ref{fig:results_measurements} (left), with a p-value of 0.001 according to the Wilcoxon signed-rank test. Similarly, TGD improved significantly, increasing from a pathological mean of $3.6\,\text{mm}$ to a pseudo healthy mean of $5.2\,\text{mm}$, as seen in Figure~\ref{fig:results_measurements} (middle), with a p-value of 0.0002.

\subsection{Evaluation of image quality}
We analyze the fastMRI test set in terms of MSE, PSNR and SSIM scores (302 volumes). All scores are calculated over the masked regions of the images only. Our model achieved an MSE of $0.0250 \pm 0.0091$, a PSNR of $16.3449 \pm 1.9196$ and an SSIM of $0.6524 \pm 0.0651$.

\newpage
\section{Conclusion}\label{sec13}
The generated pseudo-healthy inpainting shows great promise in assisting surgeons with more informed decision-making during the planning process of trochleoplasty. The generated inpaintings show reduced signs of TD and support surgeons in the decision process of the groove design. The results demonstrate that it is possible to generate pseudo-healthy 3D MR images for TD that fit to the individual patient's patella. It is worth noticing that our generated inpaintings may not represent the perfect outcome as our training database from fastMRI may be biased in several directions.

The method has limitations in terms of patient applicability. In order for it to work, the MR image of the pathological patient should be free of excessive swelling, with the patella not dislocated. Additionally, an excessive growth plate and or a focal periphyseal edema zone may introduce bright bone characteristics not present in the training dataset, potentially reducing result quality. When applicable, the method could provide surgeons with valuable guidance. 

For the future, we plan to tackle these problems with a broader database, specifically including young patients (as the demographic MR scans from patients with TD are skewed towards younger individuals) and make use of extensive data augmentation. Moreover, allowing surgeons to manually manipulate images in, for example, Virtual Reality, such as repositioning dislocated patellae, could expand the applicability of the method to more complex cases.

In conclusion, we propose a workflow for patient-specific pseudo-healthy knee MR image generation based on WDM. The generated images could serve as a valuable guide and support preoperative decision-making, complementing surgical expertise. Our method offers the potential to enhance future planning, paving the way for more accurate and patient-tailored surgery procedures.

\backmatter

%\bmhead{Supplementary information}

%If your article has accompanying supplementary file/s please state so here. 
%Authors reporting data from electrophoretic gels and blots should supply the full unprocessed scans for key as part of their Supplementary information. This may be requested by the editorial team/s if it is missing.

%Please refer to Journal-level guidance for any specific requirements.

\bmhead{Acknowledgements}

This work was financially supported by the Werner Siemens Foundation through the MIRACLE II project.

%\newpage
\section*{Declarations}

\textbf{Conflict of interest} The authors have no competing interests to declare that are relevant to the content of this article.

\newpage
\bibliography{references}% common bib file


\begin{thebibliography}{26}
\ifx \bisbn   \undefined \def \bisbn  #1{ISBN #1}\fi
\ifx \binits  \undefined \def \binits#1{#1}\fi
\ifx \bauthor  \undefined \def \bauthor#1{#1}\fi
\ifx \batitle  \undefined \def \batitle#1{#1}\fi
\ifx \bjtitle  \undefined \def \bjtitle#1{#1}\fi
\ifx \bvolume  \undefined \def \bvolume#1{\textbf{#1}}\fi
\ifx \byear  \undefined \def \byear#1{#1}\fi
\ifx \bissue  \undefined \def \bissue#1{#1}\fi
\ifx \bfpage  \undefined \def \bfpage#1{#1}\fi
\ifx \blpage  \undefined \def \blpage #1{#1}\fi
\ifx \burl  \undefined \def \burl#1{\textsf{#1}}\fi
\ifx \doiurl  \undefined \def \doiurl#1{\url{https://doi.org/#1}}\fi
\ifx \betal  \undefined \def \betal{\textit{et al.}}\fi
\ifx \binstitute  \undefined \def \binstitute#1{#1}\fi
\ifx \binstitutionaled  \undefined \def \binstitutionaled#1{#1}\fi
\ifx \bctitle  \undefined \def \bctitle#1{#1}\fi
\ifx \beditor  \undefined \def \beditor#1{#1}\fi
\ifx \bpublisher  \undefined \def \bpublisher#1{#1}\fi
\ifx \bbtitle  \undefined \def \bbtitle#1{#1}\fi
\ifx \bedition  \undefined \def \bedition#1{#1}\fi
\ifx \bseriesno  \undefined \def \bseriesno#1{#1}\fi
\ifx \blocation  \undefined \def \blocation#1{#1}\fi
\ifx \bsertitle  \undefined \def \bsertitle#1{#1}\fi
\ifx \bsnm \undefined \def \bsnm#1{#1}\fi
\ifx \bsuffix \undefined \def \bsuffix#1{#1}\fi
\ifx \bparticle \undefined \def \bparticle#1{#1}\fi
\ifx \barticle \undefined \def \barticle#1{#1}\fi
\bibcommenthead
\ifx \bconfdate \undefined \def \bconfdate #1{#1}\fi
\ifx \botherref \undefined \def \botherref #1{#1}\fi
\ifx \url \undefined \def \url#1{\textsf{#1}}\fi
\ifx \bchapter \undefined \def \bchapter#1{#1}\fi
\ifx \bbook \undefined \def \bbook#1{#1}\fi
\ifx \bcomment \undefined \def \bcomment#1{#1}\fi
\ifx \oauthor \undefined \def \oauthor#1{#1}\fi
\ifx \citeauthoryear \undefined \def \citeauthoryear#1{#1}\fi
\ifx \endbibitem  \undefined \def \endbibitem {}\fi
\ifx \bconflocation  \undefined \def \bconflocation#1{#1}\fi
\ifx \arxivurl  \undefined \def \arxivurl#1{\textsf{#1}}\fi
\csname PreBibitemsHook\endcsname

%%% 1
\bibitem[\protect\citeauthoryear{Batailler and Neyret}{2018}]{batailler_trochlear_2018}
\begin{barticle}
\bauthor{\bsnm{Batailler}, \binits{C.}},
\bauthor{\bsnm{Neyret}, \binits{P.}}:
\batitle{Trochlear dysplasia: imaging and treatment options}.
\bjtitle{EFORT open reviews}
\bvolume{3}(\bissue{5}),
\bfpage{240}--\blpage{247}
(\byear{2018})
\end{barticle}
\endbibitem

%%% 2
\bibitem[\protect\citeauthoryear{Bollier and Fulkerson}{2011}]{bollier_role_2011}
\begin{barticle}
\bauthor{\bsnm{Bollier}, \binits{M.}},
\bauthor{\bsnm{Fulkerson}, \binits{J.P.}}:
\batitle{The role of trochlear dysplasia in patellofemoral instability}.
\bjtitle{JAAOS-Journal of the American Academy of Orthopaedic Surgeons}
\bvolume{19}(\bissue{1}),
\bfpage{8}--\blpage{16}
(\byear{2011})
\end{barticle}
\endbibitem

%%% 3
\bibitem[\protect\citeauthoryear{Liebensteiner et~al.}{}]{liebensteiner_diagnostik_2023}
\begin{botherref}
\oauthor{\bsnm{Liebensteiner}, \binits{M.}},
\oauthor{\bsnm{Dirisamer}, \binits{F.}}, et al.:
Diagnostik des patellofemoralgelenks.
Arthroskopie
\textbf{36}(6),
371--372.
Accessed 2024-03-06
\end{botherref}
\endbibitem

%%% 4
\bibitem[\protect\citeauthoryear{Dirisamer et~al.}{2023}]{dirisamer_klinische_2023}
\begin{barticle}
\bauthor{\bsnm{Dirisamer}, \binits{F.}},
\bauthor{\bsnm{Attal}, \binits{R.}}, \betal:
\batitle{Klinische untersuchung des patellofemoralgelenks}.
\bjtitle{Arthroskopie}
\bvolume{36}(\bissue{6}),
\bfpage{387}--\blpage{401}
(\byear{2023})
\end{barticle}
\endbibitem

%%% 5
\bibitem[\protect\citeauthoryear{Wierer et~al.}{2023}]{wierer_therapiealgorithmus_2023}
\begin{barticle}
\bauthor{\bsnm{Wierer}, \binits{G.}},
\bauthor{\bsnm{Pfeiffer}, \binits{T.}}, \betal:
\batitle{Therapiealgorithmus der patellainstabilit{\"a}t}.
\bjtitle{Arthroskopie}
\bvolume{36}(\bissue{6}),
\bfpage{415}--\blpage{418}
(\byear{2023})
\end{barticle}
\endbibitem

%%% 6
\bibitem[\protect\citeauthoryear{Hasler and Studer}{2016}]{hasler_patella_2016}
\begin{barticle}
\bauthor{\bsnm{Hasler}, \binits{C.C.}},
\bauthor{\bsnm{Studer}, \binits{D.}}:
\batitle{Patella instability in children and adolescents}.
\bjtitle{EFORT open reviews}
\bvolume{1}(\bissue{5}),
\bfpage{160}--\blpage{166}
(\byear{2016})
\end{barticle}
\endbibitem

%%% 7
\bibitem[\protect\citeauthoryear{Fithian et~al.}{2004}]{fithian_epidemiology_2004}
\begin{barticle}
\bauthor{\bsnm{Fithian}, \binits{D.C.}},
\bauthor{\bsnm{Paxton}, \binits{E.W.}}, \betal:
\batitle{Epidemiology and natural history of acute patellar dislocation}.
\bjtitle{The American journal of sports medicine}
\bvolume{32}(\bissue{5}),
\bfpage{1114}--\blpage{1121}
(\byear{2004})
\end{barticle}
\endbibitem

%%% 8
\bibitem[\protect\citeauthoryear{Bl{\o}nd and Barfod}{2023}]{blond_trochlear_2023}
\begin{barticle}
\bauthor{\bsnm{Bl{\o}nd}, \binits{L.}},
\bauthor{\bsnm{Barfod}, \binits{K.W.}}:
\batitle{Trochlear shape and patient-reported outcomes after arthroscopic deepening trochleoplasty and medial patellofemoral ligament reconstruction: a retrospective cohort study including mri assessments of the trochlear groove}.
\bjtitle{Orthopaedic Journal of Sports Medicine}
\bvolume{11}(\bissue{5}),
\bfpage{23259671231171378}
(\byear{2023})
\end{barticle}
\endbibitem

%%% 9
\bibitem[\protect\citeauthoryear{Bl{\o}nd}{2015}]{blond_arthroscopic_2015}
\begin{barticle}
\bauthor{\bsnm{Bl{\o}nd}, \binits{L.}}:
\batitle{Arthroscopic deepening trochleoplasty: the technique}.
\bjtitle{Operative Techniques in Sports Medicine}
\bvolume{23}(\bissue{2}),
\bfpage{136}--\blpage{142}
(\byear{2015})
\end{barticle}
\endbibitem

%%% 10
\bibitem[\protect\citeauthoryear{Dejour et~al.}{1994}]{dejour_factors_1994}
\begin{barticle}
\bauthor{\bsnm{Dejour}, \binits{H.}},
\bauthor{\bsnm{Walch}, \binits{G.}}, \betal:
\batitle{Factors of patellar instability: an anatomic radiographic study}.
\bjtitle{Knee Surgery, Sports Traumatology, Arthroscopy}
\bvolume{2},
\bfpage{19}--\blpage{26}
(\byear{1994})
\end{barticle}
\endbibitem

%%% 11
\bibitem[\protect\citeauthoryear{Biedert and Albrecht}{2006}]{biedert_patellotrochlear_2006}
\begin{barticle}
\bauthor{\bsnm{Biedert}, \binits{R.M.}},
\bauthor{\bsnm{Albrecht}, \binits{S.}}:
\batitle{The patellotrochlear index: a new index for assessing patellar height}.
\bjtitle{Knee Surgery, Sports Traumatology, Arthroscopy}
\bvolume{14},
\bfpage{707}--\blpage{712}
(\byear{2006})
\end{barticle}
\endbibitem

%%% 12
\bibitem[\protect\citeauthoryear{Hinckel et~al.}{2015}]{hinckel_trochleoplasty_2015}
\begin{barticle}
\bauthor{\bsnm{Hinckel}, \binits{B.B.}},
\bauthor{\bsnm{Arendt}, \binits{E.A.}}, \betal:
\batitle{Trochleoplasty: historical overview and dejour technique}.
\bjtitle{Operative Techniques in Sports Medicine}
\bvolume{23}(\bissue{2}),
\bfpage{114}--\blpage{122}
(\byear{2015})
\end{barticle}
\endbibitem

%%% 13
\bibitem[\protect\citeauthoryear{Nolan et~al.}{2018}]{nolan_trochleoplasty_2018}
\begin{barticle}
\bauthor{\bsnm{Nolan}, \binits{J.E.}},
\bauthor{\bsnm{Schottel}, \binits{P.C.}}, \betal:
\batitle{Trochleoplasty: indications and technique}.
\bjtitle{Current Reviews in Musculoskeletal Medicine}
\bvolume{11},
\bfpage{231}--\blpage{240}
(\byear{2018})
\end{barticle}
\endbibitem

%%% 14
\bibitem[\protect\citeauthoryear{Von~Knoch et~al.}{2006}]{knoch_trochleaplasty_2006}
\begin{barticle}
\bauthor{\bsnm{Von~Knoch}, \binits{F.}},
\bauthor{\bsnm{B{\"o}hm}, \binits{T.}}, \betal:
\batitle{Trochleaplasty for recurrent patellar dislocation in association with trochlear dysplasia: a 4-to 14-year follow-up study}.
\bjtitle{The Journal of Bone \& Joint Surgery British Volume}
\bvolume{88}(\bissue{10}),
\bfpage{1331}--\blpage{1335}
(\byear{2006})
\end{barticle}
\endbibitem

%%% 15
\bibitem[\protect\citeauthoryear{Barbosa et~al.}{2024}]{barbosa_knee_2024}
\begin{botherref}
\oauthor{\bsnm{Barbosa}, \binits{R.M.}},
\oauthor{\bsnm{Serrador}, \binits{L.}}, et al.:
Knee landmarks detection via deep learning for automatic imaging evaluation of trochlear dysplasia and patellar height.
European Radiology,
1--12
(2024)
\end{botherref}
\endbibitem

%%% 16
\bibitem[\protect\citeauthoryear{Fang et~al.}{2024}]{fang_patient-specific_2024}
\begin{botherref}
\oauthor{\bsnm{Fang}, \binits{X.}},
\oauthor{\bsnm{Deng}, \binits{H.H.}}, et al.:
Patient-specific reference model estimation for orthognathic surgical planning.
International Journal of Computer Assisted Radiology and Surgery,
1--9
(2024)
\end{botherref}
\endbibitem

%%% 17
\bibitem[\protect\citeauthoryear{Van~Haver et~al.}{2014}]{van_haver_statistical_2014}
\begin{barticle}
\bauthor{\bsnm{Van~Haver}, \binits{A.}},
\bauthor{\bsnm{Mahieu}, \binits{P.}}, \betal:
\batitle{A statistical shape model of trochlear dysplasia of the knee}.
\bjtitle{The Knee}
\bvolume{21}(\bissue{2}),
\bfpage{518}--\blpage{523}
(\byear{2014})
\end{barticle}
\endbibitem

%%% 18
\bibitem[\protect\citeauthoryear{Beaufils et~al.}{2012}]{beaufils_trochleoplasty_2012}
\begin{barticle}
\bauthor{\bsnm{Beaufils}, \binits{P.}},
\bauthor{\bsnm{Thaunat}, \binits{M.}}, \betal:
\batitle{Trochleoplasty in major trochlear dysplasia: current concepts}.
\bjtitle{Sports Medicine, Arthroscopy, Rehabilitation, Therapy \& Technology}
\bvolume{4},
\bfpage{1}--\blpage{8}
(\byear{2012})
\end{barticle}
\endbibitem

%%% 19
\bibitem[\protect\citeauthoryear{Durrer et~al.}{2024}]{durrer_denoising_2024}
\begin{bchapter}
\bauthor{\bsnm{Durrer}, \binits{A.}}, \betal:
\bctitle{Denoising diffusion models for 3d healthy brain tissue inpainting}.
In: \bbtitle{MICCAI Workshop on Deep Generative Models},
pp. \bfpage{87}--\blpage{97}
(\byear{2024}).
\bcomment{Springer}
\end{bchapter}
\endbibitem

%%% 20
\bibitem[\protect\citeauthoryear{Friedrich et~al.}{2024}]{friedrich_wdm_2024}
\begin{bchapter}
\bauthor{\bsnm{Friedrich}, \binits{P.}}, \betal:
\bctitle{Wdm: 3d wavelet diffusion models for high-resolution medical image synthesis}.
In: \bbtitle{MICCAI Workshop on Deep Generative Models},
pp. \bfpage{11}--\blpage{21}
(\byear{2024}).
\bcomment{Springer}
\end{bchapter}
\endbibitem

%%% 21
\bibitem[\protect\citeauthoryear{Gu et~al.}{2024}]{gu_segmentanybone_2024}
\begin{botherref}
\oauthor{\bsnm{Gu}, \binits{H.}},
\oauthor{\bsnm{Colglazier}, \binits{R.}}, et al.:
Segmentanybone: A universal model that segments any bone at any location on mri.
arXiv preprint arXiv:2401.12974
(2024)
\end{botherref}
\endbibitem

%%% 22
\bibitem[\protect\citeauthoryear{Li et~al.}{2020}]{li_wavelet_2020}
\begin{bchapter}
\bauthor{\bsnm{Li}, \binits{Q.}},
\bauthor{\bsnm{Shen}, \binits{L.}}, \betal:
\bctitle{Wavelet integrated cnns for noise-robust image classification}.
In: \bbtitle{Proceedings of the IEEE/CVF Conference on Computer Vision and Pattern Recognition},
pp. \bfpage{7245}--\blpage{7254}
(\byear{2020})
\end{bchapter}
\endbibitem

%%% 23
\bibitem[\protect\citeauthoryear{Knoll et~al.}{2020}]{noauthor_fastmri_nodate}
\begin{barticle}
\bauthor{\bsnm{Knoll}, \binits{F.}},
\bauthor{\bsnm{Zbontar}, \binits{J.}}, \betal:
\batitle{fastmri: A publicly available raw k-space and dicom dataset of knee images for accelerated mr image reconstruction using machine learning}.
\bjtitle{Radiology: Artificial Intelligence}
\bvolume{2}(\bissue{1}),
\bfpage{190007}
(\byear{2020})
\end{barticle}
\endbibitem

%%% 24
\bibitem[\protect\citeauthoryear{Zbontar et~al.}{2018}]{zbontar_fastmri_2019}
\begin{botherref}
\oauthor{\bsnm{Zbontar}, \binits{J.}},
\oauthor{\bsnm{Knoll}, \binits{F.}}, et al.:
fastmri: An open dataset and benchmarks for accelerated mri.
arXiv preprint arXiv:1811.08839
(2018)
\end{botherref}
\endbibitem

%%% 25
\bibitem[\protect\citeauthoryear{Kikinis et~al.}{2013}]{kikinis_3d_2014}
\begin{bchapter}
\bauthor{\bsnm{Kikinis}, \binits{R.}},
\bauthor{\bsnm{Pieper}, \binits{S.D.}}, \betal:
\bctitle{3d slicer: a platform for subject-specific image analysis, visualization, and clinical support}.
In: \bbtitle{Intraoperative Imaging and Image-guided Therapy},
pp. \bfpage{277}--\blpage{289}.
\bpublisher{Springer},
\blocation{New York, NY, USA}
(\byear{2013})
\end{bchapter}
\endbibitem

%%% 26
\bibitem[\protect\citeauthoryear{}{}]{noauthor_3d_nodate}
\begin{botherref}
3D Slicer Image Computing Platform.
\url{https://slicer.org/}
Accessed 2024-09-11
\end{botherref}
\endbibitem

\end{thebibliography}
%\printbibliography


\end{document}